\documentclass{article}

 \PassOptionsToPackage{numbers, compress}{natbib}

\usepackage[final]{nips_2017}

\usepackage[utf8]{inputenc} % allow utf-8 input
\usepackage[T1]{fontenc}    % use 8-bit T1 fonts
\usepackage{hyperref}       % hyperlinks
\usepackage{url}            % simple URL typesetting
\usepackage{booktabs}       % professional-quality tables
\usepackage{amsfonts}       % blackboard math symbols
\usepackage{nicefrac}       % compact symbols for 1/2, etc.
\usepackage{microtype}      % microtypography

\title{BioMM: Biologically-informed Multi-stage Machine learning for identification of epigenetic fingerprints}

\author{
  Junfang Chen \\
  Central Institute of Mental Health \\
  Heidelberg University, Germany \\ 
  \texttt{junfang.chen@zi-mannheim.de} \\
    \And
  Emanuel Schwarz* \\
  Central Institute of Mental Health \\
  Heidelberg University, Germany  \\ 
  \texttt{emanuel.schwarz@zi-mannheim.de} \\
}

\begin{document}

\maketitle

\begin{abstract}
  
The identification of reproducible biological patterns from high-dimensional data is a bottleneck for understanding the biology of complex illnesses such as schizophrenia. To address this, we developed a biologically informed, multi-stage machine learning (BioMM) framework. BioMM incorporates biological pathway information to stratify and aggregate high-dimensional biological data. We demonstrate the utility of this method using genome-wide DNA methylation data and show that it substantially outperforms conventional machine learning approaches. Therefore, the BioMM framework may be a fruitful machine learning strategy in high-dimensional data and be the basis for future, integrative analysis approaches.

\end{abstract}

\section{Introduction}

The identification of predictive biological signatures from high-dimensional biological data remains a major challenge. This particularly applies for clinically and biologically heterogeneous illnesses, such as schizophrenia \cite{mcgrath2008schizophrenia}. Schizophrenia is defined based on clinical symptom constellations, is highly polygenic and is thought to result from complex gene-environment interactions. As a consequence, effect sizes of individual biological markers is generally low, creating an opportunity for machine learning approaches to identify more predictive biological fingerprints. Polygenic scores integrating common variation across the genome explain a substantial amount of heritable variance, but are not sufficiently predictive for diagnostic applications \cite{schizophrenia2014biological, network2015psychiatric}. Due to the importance of environmental risk \cite{cantor2005schizophrenia, swofford1996substance, brown2009prenatal, teague1989stress}, analysis of epigenetic modifications, such as DNA methylation \cite{grayson2013dynamics}, has received considerable attention in psychiatric research \cite{nishioka2012dna, roth2009epigenetic}. The analysis of such epigenetic data entails several challenges, including its sensitivity to confounding factors such as medication, which have thus far limited the utility of this modality for integrative analytics in psychiatry. Due to the high data dimensionality and frequently non-reproducible association signals, it remains to explored whether schizophrenia is characterized by consistent patterns of abnormal DNA methylation. 
Here, we aimed to develop a novel strategy to derive reproducible biological patterns from DNA methylation data. Our approach builds on the hypothesis that biological stratification can meaningfully reduce the dimensionality of the data. Specifically, inspired by functional genomics studies \cite{pan2014system, liu2017pathway}, we built a biologically informed two-stage learning framework. At the first level, machine learning is performed independently on biological predictors (i.e. DNA methylation probes) harboured by gene sets of different ontological categories. This is based on the hypothesis that these gene sets capture at least partially independent, illness associated methylation effects. The predicted scores were then used as input for machine learning at the second level, to identify a predictive methylation pattern across ontological categories. Testing of the algorithm was performed in an independent test dataset, to capture its robustness against cross-dataset heterogeneity. To evaluate the utility of this strategy, we performed a comparative analysis against 5 conventional machine learning approaches. We demonstrate that the biologically informed, multi-stage machine learning significantly outperforms these conventional approaches.

\section{Materials and Methods}
\subsection{Genome-wide methylation data}

Genome-wide profiles of DNA methylation were obtained from two independent cohorts with a total sample number of 1522 (\textbf{Table~\ref{demographicTable}}). Data were downloaded from the GEO database \cite{edgar2002gene}. Detailed descriptions of cohorts and data acquisition can be found elsewhere \cite{hannon2016integrated}. We focussed on the overlapping set of autosomal methylation sites to limit the potential influence of sex on machine learning due to the phenomenon of X chromosome inactivation or the existence of an additional X chromosome in female samples. The dataset GSE80417 was used as training set and GSE84727 for testing (independent test set) of machine learning algorithms.

\begin{table}
\centering
\caption{Overview of demographic information}
	\label{demographicTable}
\begin{tabular}{ | p{5cm} | p{1.1cm} | p{1.1cm} | p{1.5cm} | p{1.5cm} |r|}
	
  \hline
  \centering \textbf{  Meta-information} & \centering  \textbf{Controls} & \centering \textbf{Cases} & \centering  \textbf{Sex (m/f)} &  \textbf{  Age } \\ \hline
    \centering GEO:GSE80417 (phase 1 cohort)  & \centering 322 & \centering 353 & \centering 396/279 &  ${40.5}\pm${15.2} \\ \hline 
    \centering GEO:GSE84727 (phase 2 cohort)  & \centering 433 & \centering 414	& \centering 602/245 &  ${44.6}\pm${12.9}  \\  \hline

 \end{tabular} 
\end{table} 

\subsection{Adjustment for potential confounders}
The data was corrected to account for the influence of potential confounders, which comprised cigarette smoking \cite{de2005meta}, population structure \cite{liu2010identification}, cellular composition, gender and age. Smoking was quantified from DNA methylation levels as described previously \cite{elliott2014differences, zeilinger2013tobacco}. Population structure was determined from methylation data via Principal Components Analysis. Specifically, the first 10 principal components were considered as covariates. Cellular composition was quantified using the Epigenetic Clock tool (\url{https://dnamage.genetics.ucla.edu/}) \cite{horvath2013dna} and included the seven recommended cell types: CD8.naive, CD8pCD28nCD45RAn, PlasmaBlast, CD4T, NK, Mono, Gran. All covariates were used in a linear model to residualize each given methylation probe. This was performed separately for both cohorts and the resulting residuals were used for downstream analysis.

\subsection{Gene and ontological assignment}
For each gene, we extracted all CpGs within 20Kb upstream and downstream of the transcription start and end sites, respectively. CpG locations were downloaded from the NCBI database for both datasets and gene boundaries from the R library TxDb.Hsapiens.UCSC.hg19.knownGene. Information on gene ontologies was derived from the R library org.Hs.eg.db. In total, 2135 ontological categories (biological processes only), each containing between 10 and 200 genes were used for analysis. 

\subsection{BioMM algorithm}

Input:
\begin{enumerate}
\item	 K: the total number of gene ontological categories (GO).
\item	 $D_{k}$: the stage-1 training set $D_{k} = \{((x_{1}^k, ...x_{M}^k), Y_{D})\}$ with M samples for every k = 1, ..., K and a fixed label $Y_{D}$.
\item	 $R_{k}$: the stage-1 independent test set $R_{k} = \{((x_{1}^k, ...x_{N}^k), Y_{R})\}$ with N samples for every k = 1, ..., K and a fixed label $Y_{R}$.
\item	$f^1$: the stage-1 model. 
\item $f^2$: the stage-2 model. The random forest algorithm was used in the present study.
\end{enumerate}
 
Output: 
\begin{enumerate}
\item	 $\hat{Y}_{kM}^1$: the stage-1 bootstrapping prediction score with M samples for $k^{th}$ GO.
\item	 $\hat{Y}_{kN}^1$: the stage-1 independent test prediction score with N samples for $k^{th}$ GO.
\item	 $D^2$: the stage-2 training set $D^2= {(\hat{Y}_{1M}^1,…, \hat{Y}_{KM}^1, Y_{D})}$ with M samples and the label $Y_{D}$.
\item  $R^2$: the stage-2 test  set $R^2= {(\hat{Y}_{1N}^1,…, \hat{Y}_{KN}^1, Y_{R})}$ with N samples and the label $Y_{R}$.
\item	 $\hat{Y}_{test}^2$: the stage-2 test estimate. 
\item	 $Err_{test}$: the error rate for the stage-2 test estimate. $Err_{test} = L(Y_{R}, \hat{Y}_{test}^2)$ with the 0-1 loss function L.
\end{enumerate}

\underline{BioMM 1\textsuperscript{st} stage:}  

\begin{enumerate}
	\item For each ontological category $k$, repeat the following steps B times (here, B=100):
	\begin{enumerate} 
		\item Draw a bootstrap set from $D_{k}$ with the same sample size. The out-of-bag samples are used as test set.
		\item Fit a machine learning model $f^1$ to the bootstrapped data.
		\item Predict the model on the test set and $R_{k}$.  
	\end{enumerate}	 

	\item For each GO $k$, determine the averaged prediction score $\hat{Y}_{kM_{avg}}^1 = \frac{1}{B}\sum_{b=1}^{B} \hat{Y}_{kM}^1$ and $\hat{Y}_{kN_{avg}}^1 = \frac{1}{B}\sum_{b=1}^{B} \hat{Y}_{kN}^1$.
	\item Combine $\hat{Y}_{kM_{avg}}^1$ and $\hat{Y}_{kN_{avg}}^1$ across all $K$ GOs to generate $D^2$ and $R^2$.
 \end{enumerate}
  
\underline{BioMM 2\textsuperscript{nd} stage:}  

\begin{enumerate}
\item	Fit a machine learning model  $f^2$ to $D^2$ with features that are positively correlated with $Y_{D}$.
\item	Predict the model on $R^2$ with the corresponding feature set. 
\item	Repeat steps 1 and 2 20 times and average prediction scores to obtain $\hat{Y}_{test}^2$. 
\item	Determine $Err_{test}$ to evaluate the model performance. 
 \end{enumerate}

\subsection{Classifier selection}
To compare classifiers regarding performance, five different well-known classifiers were used as machine learning models for the first stage of BioMM: random forests, support vector machine (SVM) with the radial basis function kernel, logit regression, the LASSO and elastic net. No variable selection was performed since the categorization based on ontological categories already substantially reduced data dimensionality. At the second stage, the random forest algorithm was applied to capture potential interactions between ontological categories. 

\subsection{Comparison against conventional machine learning}
The same five classifiers were used as conventional machine learning tools to compare performance against the BioMM method. Due to the large data dimensionality, variable selection was performed based on Pearson correlation filtering. The number of selected features (5, 10, 20, 30, 50, 100, 500, 1000 or 3000 CpGs) was determined using bootstrapping in the training data.

\section{Results and Discussion}

\begin{table}
\centering
\caption{Classification performance (quantified as error rates) of conventional classifiers and BioMM}
	\label{ClassificationPerformance}
\begin{tabular}{ | p{3.3cm} | p{2.4cm} | p{1.1cm} | p{1.1cm} | p{1.4cm} | p{1.1cm} |r|}
	 
  \hline
  \centering \textbf{Classifiers} & \centering  \textbf{Random forest} & \centering \textbf{SVM} & \centering  \textbf{logit} & \centering  \textbf{  LASSO }  &  \textbf{Elastic} \\ \hline 
    \centering \textbf{Conventional\_$Err_{test}$} & \centering 0.511 & \centering 0.519 & \centering 0.514	&  \centering  0.515 &  0.514 \\ \hline 
    \centering \textbf{BioMM\_$Err_{test}$}	     & \centering \textbf{0.391} & \centering 0.432	& \centering 0.482 & \centering   0.509 &  0.502  \\  \hline
 \end{tabular} 
\end{table}

\textbf{Table~\ref{ClassificationPerformance}} shows that BioMM using random forests outperformed conventional and other BioMM approaches, with a classification error rate of 39.1\%. The second best classifier was BioMM using SVM. The superior performance of BioMM using random forests may have been due to the following reasons: (I) Random forests can better capture interactions between predictors, compared to the remaining models. Notably, application of conventional random forests was not able to yield comparative performance, likely due to the fact that variable selection was also based on a linear method. Similarly, non-linear SVM using radial basis functions, showed substantially better performance as part of the BioMM, compared to conventional machine learning approach. The possible advantage of capturing predictor interactions would relate to interactions of individual methylation sites only, since there was no pathway-level aggregation for conventional machine learning and all second stage aggregation part of BioMM applications were performed using random forests. (II) An interesting aspect of the BioMM method is that, in contrast to conventional machine learning, there is redundancy in the predictor set that may impact positively on classification. The redundancy arising from gene overlap between ontological categories creates the possibility for the same methylation probes to be part of different rules, potentially exploiting better additive and multiplicative effects \cite{takahashi2010diagnostic}. The ontological stratification of methylation sites may yield second-stage predictors that are independently associated with outcome and therefore lead to an increase in classification performance. These independent effects may be lost in non-stratified data due to its high dimensionality. (III) Repeated sampling at the first and second stage can average over sampling variability and lead to predictions with higher biological reproducibility (1\textsuperscript{st} stage) and accuracy (2\textsuperscript{nd} stage). (IV) Illness associated methylation differences may be hierarchically organized such that alterations of individual methylation sites impact first at the pathway level which subsequently leads to a systems-wide effect.  
\\
The BioMM method conceptually aligns well with the growing consensus that schizophrenia is not a single disease entity, but composed of multiple subtypes with different biological underpinnings. In contrast to conventional machine learning (as well as the variable selection performed here), which generally builds on the assumption that all subjects originate from the same population, BioMM adds a dimension of biological stratification. It specifically exploits the hypothesis that some subjects may have alterations in a given biological process (thereby receiving high prediction scores at the first stage) and these subgroups of patients are then integrated at the second stage. Therefore, exploring how first stage predictors distribute across the patient cohorts may give novel insights into the presence of potential patient subgroups. This may allow a personalized extension functional analysis, which had previously been performed on synthetic features similar to those created by BioMM after first stage computations \cite{pan2014system, liu2017pathway}. Notably, these features do not allocate subjects to mutually exclusive clusters but may elucidate overlapping patient communities depending on the implicated biological processes. In the present study, we have corrected data for several potential confounding effects that are known to modify DNA methylation levels. However, we cannot exclude that other factors, such as medication or lifestyle effects, may have introduced bias and led to artifactual illness-associated profiles. Therefore, the present results should be replicated in additional cohorts that are not affected by these confounders. Results from conventional machine learning, as well as the BioMM method using some first-stage classifiers, however, demonstrate that these effects did not lead to obvious signatures that were reproducible across datasets. The resulting signatures may, therefore, hint at the existence of reproducible methylation signatures in schizophrenia and show utility for integrative analyses with other data modalities. 
\\
From a computational perspective, further improvements to the method may be achieved by variable selection during the first and second stage. At the second stage, we selected predictors positively associated with outcome in the training data to prevent confounding, but more advanced variable selection procedures may further improve performance \cite{perlich2011cross}. Additionally, especially if larger data sets are available, other strategies of creating ‘synthetic’ data sets during the first stage may be preferable. Here, we used the oob predictions, but these have been reported to be conservatively biased \cite{mitchell2011bias}. This could impact on the comparability between the synthetic training and test data, which could, in turn, negatively impact on classifier performance. Another potential concern is that given the high number of ontological categories used for stratification, some synthetic features will likely show illness associations by chance and, therefore, receive high classifier weights at the second level. This is particularly a problem if the true biological signal is weak and distributed across numerous ontological categories. Besides additional data to remove false-positive associations, other data modalities or biological meta-information, such as tissue specific expression information, may help to focus on biologically relevant signals. Finally, we expect the BioMM strategy to be a fruitful approach for integration of multiple data modalities that can be mapped via genetic and ontological information, such as with genetic association or expression data.

\section{Conclusion}

We have developed here a biologically informed machine learning framework that aims to identify reproducible biological patterns through biological stratification and aggregation of high-dimensional data. This BioMM method outperformed conventional machine learning approaches based on evaluation of prediction in independent test data. This computational framework may allow the exploration of patient subgroup effects and show utility for integrative analyses with other data modalities.

\subsubsection*{Acknowledgments}
 
We thank Dr Hannah Elliott (University of Bristol MRC Integrative Epidemiology Unit) for providing code to calculate DNA methylation smoking scores. This study was supported by the Deutsche Forschungsgemeinschaft (DFG), SCHW 1768/1-1.

\bibliography{bib}
\bibliographystyle{plain} 

\end{document}